# Ransomware and Artificial Intelligence: A Comprehensive Systematic Review of Reviews

**Therdpong Daengsi[1], Phisit Pornpongtechavanich[2], Paradorn Boonpoor[1], Kathawut Wattanachukul[1,5], Korn Puangnak[3], Kritphon Phanrattanachai[4], Pongpisit Wuttidittachotti[5], Paramate Horkaew[6]**

[1]Department of Sustainable Industrial Management Engineering, Faculty of Engineering, Rajamangala University of Technology Phra Nakhon, Bangkok, Thailand
[2]Department of Information Technology, Faculty of Industry and Technology, Rajamangala University of Technology Rattanakosin Wang Klai Kangwon Campus, Prachuap Khiri Khan, Thailand
[3]Department of Computer Engineering, Faculty of Engineering, Rajamangala University of Technology Phra Nakhon, Bangkok, Thailand
[4]Department of Electronics and Automation Systems Engineering, Faculty of Agricultural and Industrial Technology, Phetchabun Rajabhat University, Phetchabun, Thailand
[5]Department of Digital Network and Information Security Management, Faculty of Information Technology and Digital Innovation, King Mongkut's University of Technology North Bangkok, Bangkok, Thailand
[6]Department of Computer Engineering, Institute of Engineering, Suranaree University of Technology, Nakhon Ratchasima, Thailand

## Article Info

## ABSTRACT

This study provides a comprehensive synthesis of Artificial Intelligence (AI), especially Machine Learning (ML) and Deep Learning (DL) in ransomware defense. Using a "review of reviews" methodology based on the PRISMA, this paper gathers insights on how AI is transforming ransomware detection, prevention, and mitigation strategies in the past five years (2020-2024). The findings highlight the effectiveness of hybrid models, which combine multiple analysis techniques such as code inspection (static analysis) and behavior monitoring during execution (dynamic analysis). The study also explores anomaly detection and early warning mechanisms before encryption that tackle ransomware's growing complexity. It also examines key challenges in ransomware defense, such as techniques designed to deceive AI driven detection, and the lack of strong and diverse datasets. It highlights AI's role in early detection and real-time response systems, enhancing scalability and resilience. With the systematic review of reviews approach, the contributions of this study are systematically consolidating research insights from multiple review articles, identifying effective AI models, and bridging theory with practice to foster collaboration among academia, industry, and policymakers. Future research directions are anticipated and practical recommendations for cybersecurity practitioners are provided. Finally, it presents a roadmap for advancing AI-driven countermeasures, for the protection of key systems and infrastructures against evolving ransomware threats.

*This is an open access article under the CC BY-SA license.*

*Corresponding Author:*

Phisit Pornpongtechavanich
Department of Information Technology, Faculty of Industry and Technology, Rajamangala University of Technology Rattanakosin Wang Klai Kangwon Campus, Prachuap Khiri Khan, Thailand
Phet Kasem Rd., Nong Kae, Hua Hin District, Prachuap Khiri Khan 77110
Email: phisit.kha@rmutr.ac.th

## 1. INTRODUCTION

Ransomware is a type of malicious software that has become a most severe cyber threats in today's digital world. It has significantly impacted businesses, economic entities, governments, public infrastructure, and individuals alike, often causing major disruptions to operations and substantial financial losses [1], [2]. The situation is escalating, with data in 2023 showing that approximately 73% of businesses globally reporting ransomware incident [3]. It is the highest rate seen between 2018 and 2023 [3]. In that year, the US healthcare sector was severely impacted by the attacks [4]. As a result, it suffered the highest number of data breaches. According to the re-ports, critical manufacturing ranked as the second most targeted sector, followed closely by government entities and their essential infrastructures [4].

This malicious software works by locking or encrypting victim's data, making it impossible to access until a ransom demand is met [5]. Currently, payments are usually extorted in cryptocurrencies such as Bitcoin [6], [7]. This is because they are difficult to trace thanks to their TOR systems [1]. These new currencies help cybercriminals evade detection and tracking, especially when using cryptocurrency integrated process [8]. Ransomware has recently evolved and rapidly become more sophisticated. It targets not only individuals but also organizations, including financial institutions [8], [9], hospitals [10], [11], transportation systems [12], industrial facilities [13], energy, [14] and critical infrastructures [1], [13], as well as government agencies and services [14]. These attacks have caused major disruptions and financial losses worldwide [9], [13]. Technological advancements (e.g., the emergence of the Internet of Things (IoT) and Industry 4.0) have made these attacks even more frequent [13], [15]. IoT connects computing devices to the internet, while Industry 4.0 auto-mates manufacturing processes and supply chains. While being useful, these innovations also introduce new secu-rity risks [13]. Cybercriminals often seek and exploit vulnerabilities, weaknesses or improperly configurations in these systems to launch ransomware attacks [16], [17]. The emergence of Ransomware-as-a-Service (RaaS) has played a significant role in the escalation of ransomware incidents by providing easy-to-modify ransomware kits and easy-to-use tools [18], [19], thereby enabling even non-technical actors or everyone to efficiently organize and execute attacks [20].

The financial sector is particularly vulnerable to ransomware [9]. Banks and payment systems rely on real-time data, making them prime targets [21]. Criminals often use phishing emails to breach these systems, aiming to steal data or disrupt services [22]. Cybersecurity experts are increasingly interested in artificial intelligence (AI) for its ability to quickly analyze large amounts of internet traffic and detect cyberattacks [23]. As a result, financial institutions are turning to AI and Machine Learning (ML), with expert support, to enhance their defenses [8], [24]. These tools can scan huge datasets to spot unusual patterns or threats. However, using AI and ML also presents challenges, especially regarding regulations and customer privacy [2], [8]. The manufacturing sector faces similar ransomware threats [13]. Automated systems need a constant flow of data to operate. A cyberattack could halt production lines, leading to significant losses [13]. The complex nature of these networks also makes attacks harder to detect and stop.

Ransomware has severely impacted the healthcare sector [10], [11], [25]. Hospitals need instant access to patient records for good care. If these systems are locked, treatments can slow down, potentially risking lives [10], [11]. Like other victims, cybercriminals target healthcare providers because they're likely to pay quickly to get essential data back. These attacks highlight the need for strong security to protect sensitive patient information and ensure safety [1]. The COVID-19 pandemic worsened this situation [26], [27]. With more people working from home [28], many companies used less secure networks. Cybercriminals took advantage, launching phishing campaigns to get into corporate systems. A notable attack involved ransomware hidden in a real-looking app called CovidLock, which locked users' phones and demanded Bitcoin [18]. These events show how fast attackers adapt to new chances and stress how important it is to be proactive in cybersecurity.

AI is widely used across many fields to offer smart solutions and tools, aiming to improve cybersecurity [29]. To fight ransomware effectively, organizations need a mix of strong technology, clear rules, and good management. Advanced tools like machine learning (ML) and deep learning (DL) can help find, detect, and potentially stop ransomware before it causes major harm [18], [24]. These tools watch for suspicious behaviors in real-time, spotting unusual activity that might signal an attack [24], [30]. For example, systems that monitor how software behaves can block ransomware from encrypting files if they see a suspicious action [31]. However, relying solely on technology is not enough anymore. Government and international groups must work together to create laws that make it tougher for cybercriminals to operate [1], [9]. This includes sharing information about possible threats to boost cooperation between governments. Meanwhile, attackers are also using AI to make their techniques and tools more effective [31], [32]. These technologies let them find new ways to avoid being caught, making it harder for victims to prevent or defend against cyberattacks. This is because AI tools can make attacks more diverse and complex, and malware and ransomware more advanced [33]. This leads to an ongoing "arms race" where cybercriminals constantly develop new tactics, forcing cybersecurity experts to innovate and stay ahead [34], [35].

One major reason ransomware remains so successful is its profitability. Attackers can demand ransoms, often in cryptocurrency, worth millions of dollars [26], [36]. Victims, especially in critical sectors like healthcare or public services, often feel compelled to pay [10], [11]. This financial incentive has made ransomware one of the most lucrative forms of cybercrime. To reduce its impact, organizations must adopt stronger cybersecurity measures. Employees also play a crucial role, as many attacks begin with phishing emails. Training programs can help staff recognize and avoid such threats [35]. Regular system backups are essential, as quick data restoration reduces the attackers' leverage [10], [34]. Testing these backups ensures they work when needed. In addition, governments and businesses must invest in public awareness and cybersecurity research. Collaboration between private and public sectors can lead to innovative solutions. Sharing threat intelligence and mitigation strategies helps enhance collective security. International cooperation is also necessary to close legal gaps and hold cybercriminals accountable [2], [21]. Exploring ransomware trends, particularly involving AI, is not only beneficial but increasingly essential.

Instead of a conventional literature review, this study conducted a comprehensive systematic review of reviews or surveys. Using the PRISMA framework [37], it ensures a structured synthesis of research from the past five years. The methodology consolidates insights from meta-reviews and surveys to provide a concise yet broad view of AI's role in ransomware defense. It organizes key issues, trends, and findings to give a clear overview of AI-driven mitigation. The study also identifies effective AI techniques for detecting, preventing, and mitigating ransomware, bridging theory with practice and encouraging collaboration across academia, industry, and policy. Additionally, it explores future research paths and offers practical insights for applying AI in real-world cybersecurity to protect critical systems.

After this introduction section, the background of ransomware is described in Section 2. Then, the method employed by this study is presented in Section 3. Next, Sections 4 and 5 describe the roles of AI in detection and prevention, and mitigation respectively. Sections 6 and 7 highlight the challenges in ransomware defense and research using AI and challenges and future ransomware research directions associated with AI. The last two sections, the discussion and the concluding remarks are made accordingly in Section 8 and 9.

## 2. OVERVIEW OF RANSOMWARE

### 2.1. The Evolution of Ransomware

Ransomware began in 1989 [2], however, the ransomware scene changed in the early 2010s with CryptoLocker (in 2013) [7]. It used strong encryption, making data nearly impossible to recover without a decryption key. It mainly targeted individuals and small businesses via malicious email attachments [15]. This marked a shift to profit-driven cybercrime, with attackers demanding Bitcoin for anonymous and profitable global payments [7]. In 2017, the WannaCry attack became a major turning point. It used EternalBlue, a flaw in Microsoft Windows™ created by the U.S. NSA, to quickly spread across networks [38], [39]. WannaCry affected vital systems worldwide, including hospitals, transport, and government services [25], [40]. Recent ransomware focuses on high-value victims, using double extortion and advanced tactics to avoid detection and increase damage [41].

Ransomware has become very complex harmful software and organized attacks. A key development is Ransomware-as-a-Service (RaaS) [18], [19]. This business model lets people rent or buy ransomware tools or ransomware kits from developers, making it easy for even those without tech skills to use [19], [20]. By offering easy-to-use interfaces, custom options, shared profits, and tech support, RaaS has greatly increased ransomware attacks. Prominent RaaS platforms like GandCrab, LockBit, DarkSide, REvil, Dharma, Egregor, Maze and Avaddon have shaped the ransomware world [17], [42]. For example, LockBit, Maze and REvil brought in new features like custom ransom demands and double extortion [26], [27]. In the first half of 2023, LockBit and Conti were the top two detected worldwide, at 14.0% and 9.4% respectively [43]. DarkSide's attack on Egregor and Colonial Pipeline became known for harsh double extortion, showing RaaS's global impact [27], [44]. Today, modern RaaS platforms lead the ransomware scene, letting users customize and launch attacks globally with advanced ways to avoid detection, increasing their impact. Additionally, the COVID-19 pandemic significantly increased the ransomware threat to society, as attackers exploited vulnerabilities in remote work setup and made money on pandemic induced fears [18], [26]. A notable ransomware variant, i.e., CovidLock, deliberately targeted Android users using fake work tracking applications to lock their mobile devices and demand ransoms.

### 2.2. The Categories of Ransomware

Understanding ransomware types and methods is key to strong defense. Ransomware is categorized by function, target, and spread to better grasp its forms and behaviors. Crypto ransomware, or encryption ransomware, encrypts files, making them unusable until payment. It applies strong encryption (e.g., AES or RSA), often with public-private key systems. CryptoLocker, WannaCry (which caused global chaos in 2017 [18], [19]), KeRanger, and CryptoWall are examples. These typically irreversible attacks target file types like .pdf or .zip and demand payments in cryptocurrencies such as Bitcoin [7], [19]. Locker ransomware (ScreenLocker) simply locks users out of devices by taking over system controls, like input devices, without encrypting files [45]. It can disable the entire operating system interface, forcing a ransom screen [34]. Data stays safe but is only reachable after malware removal. Victims pay to regain access [27]. Variants include Curve-Tor-Bitcoin (CTB) Locker and Winlock, which lock screens, browsers, or the Master Boot Record (MBR) [19].

Ransomware targets range from individuals to large organizations. Individual attacks often use phishing emails or bad websites, exploiting weaker systems for quick payments [19]. Ryuk and Maze ransomware aim for high-value targets like hospitals, government agencies, and corporations [27], [30]. This "Big Game Hunting" (BGH) approach disrupts key operations and demands large ransoms [27]. Organizations face financial and operational damage from advanced ransomware that encrypts, exfiltrates data, or uses double extortion [30], [45]. Ransomware also differs in how it spreads. Ransomware-as-a-Service (RaaS) lets affiliates rent ready-made ransomware kits on the dark web, like GandCrab (2018-2019) [17], [27]. WannaCry, using an Server Message Block (SMB) flaw called EternalBlue, caused a global outbreak, hitting over 230,000 systems in 150 countries [17], [26]. Similarly, Locky ransomware spreads via phishing emails with harmful macros that trick users into downloading it [10], [26]. RaaS allows wide deployment, targeting both individuals and organizations, expanding the threat [22]. These models have increased frequency and impact of global ransomware attacks [27].

Lastly, ransomware can be deployed manually or automatically. Manual deployments involve attackers using stolen RDP credentials to install ransomware directly on systems, as seen with SamSam and Pony [18], [26]. Automated ones, like those using botnets, enable large-scale infection without direct attacker involvement [27], [34]. Both methods target critical systems and organizations, focusing on widespread and efficient attacks [30], [45]. While different, both manual and automated deployments contribute equally to global ransomware attacks [34], [45].

### 2.3. How Ransomware Works

Generally, attackers first find weak spots, then deliver the harmful software through exploit kits or phishing emails [28]. Once inside, the ransomware turns off security, stays hidden, and then encrypts important data. If done right, these encryptions are almost impossible to undo [18]. Victims then get ransom demands, often with threats to release sensitive data if they do not pay. Otherwise, their vital data or private organizational information might be posted on the dark web for double extortion or destroyed, damaging their reputation [17]. The details of the ransomware workflow, as illustrated in Figure 1, are described as follows:

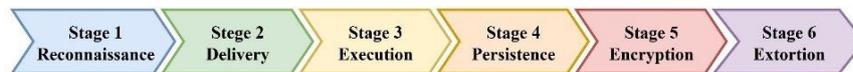

Figure 1. Six stages of the ransomware workflow

1) Reconnaissance: this stage involves gathering information about the victim's system [2]. Attackers identify vulnerabilities that can be exploited later. They include outdated software, weak passwords, or misconfigured systems, etc. Reconnaissance also includes researching potential human factors.

2) Delivery: this stage focuses on introducing the malicious payload into the target system. Common ways include phishing emails with malicious links or attachments, and automatic downloads from infected websites [2]. Once delivered, the ransomware infiltrates the system by exploiting vulnerabilities.

3) Execution: At this stage, the ransomware activates from within the infected system. It collects sensitive information about the victim's device [2]. Additionally, it may disable antivirus software, delete shadow file copies, or terminate system critical processes to prepare the environment for its primary function.

4) Persistence: To maintain control and ensure successful execution, ransomware usually establishes a persistence strategy. This may include creating registry entries, hiding in system processes, or using rootkits. Persistent ransomware has to ensure that it can resume its operations even after a system reboot [2], [27].

5) Encryption: This most critical stages, ransomware executes an encryption algorithm to encrypt files, rendering them inaccessible to the target user [10], using strong encryption algorithms (e.g., AES) are often used [17]. The ransomware may also encrypt specific file types on shared drives.

6) Extortion: Finally, an attacker usually demands a ransom on host's display [2]. A ransom note usually details the required amount and instructions on making the payment [10]. Victims are commonly required to pay in cryptocurrency to obtain the decryption key. Some variants of ransomware use double extortion tactics, threatening to publish the stolen data if the ransom demands are ignored or not reponded [26].

## 3.    METHODOLOGY

A simple search found a large number of articles related to the topic of interest [1-3], [6-36], [40-61]. Therefore, the PRISMA protocol utilized in [41], [62] was adopted to remove any bias, resulting from unsystematic and unstructured analyses of the contents using traditional review approach. The article selection steps taken in this study are illustrated in Figure 2. Each step is explained in the following subsections.

## 3.1  Step 1: Identification

Keywords and conditions were defined to search for relevant research articles. In this work, the specified keywords were "Ransomware" AND "AI" OR "Ransomware" AND "Artificial Intelligence". The conditions were then limited to articles published between 2020 and 2024. Moreover, qualified articles were obtained from IEEE Xplore, ACM, ScienceDirect, SpringerLink, Wiley, and MDPI, rather than from Scopus or Web of Science. The exclusive selection of these databases is due to access limitations imposed on the network of the authors' institutes, by certain publishers.

## 3.2  Step 2: Article Screening

The collected articles were then reviewed and linked to publication databases systematically. A total of 315 articles in the IEEE Xplore database, 633 articles in the ACM database, 1,202 articles in the ScienceDirect database, 1,796 articles in the SpringerLink database, 622 articles in the Wiley database, and 17 articles in the MDPI database were identified.

## 3.3  Step 3: Eligibility Criteria

This step involves subjecting the articles obtained from the Article Screening step to eligible conditions. Four conditions are applied:
1) Authors must have access to the full text of the article through their university network.
2) The content must be in the form of an overview, review, or survey.
3) The title, abstract, and content must be related to ransomware and artificial intelligence.
4) Each article must be published in a journal indexed in Scopus. Specifically, the journal must be ranked in Q1 or Q2 in at least one category; journals in other quartiles are excluded. After applying these four conditions, the results showed five articles from IEEE Xplore, four from ACM, seven from ScienceDirect, four from SpringerLink, two from Wiley, and four from MDPI (after applying the special exclusion criterion related to citation impact — the higher, the better), as shown in Figure 2.

However, when considering yearly publication statistics, there was no article published in 2020 that satisfied the criteria, but there were six articles published in 2021, seven articles published equally in 2022 and 2023, and eight articles published in 2024.

## 3.4  Step 4: Article Inclusion

In this step, the articles satisfying all the criteria are summarized, resulting in a total of 27 articles After obtaining the selected 27 articles, comprising 7 articles from ScienceDirect, 5 articles from IEEEXplore, 4 articles from the ACM Digital Library, 4 articles from SpringerLink, 5 articles from MDPI, and 2 articles from Wiley Online Library. The analysis by year revealed that only articles published between 2021 and 2024 met the criteria. Specifically, 6 articles were published in 2021, 7 in 2022, 7 in 2023, and 7 in 2024. No articles published in 2020 met the criteria. These 27 articles have been considered and analyzed.

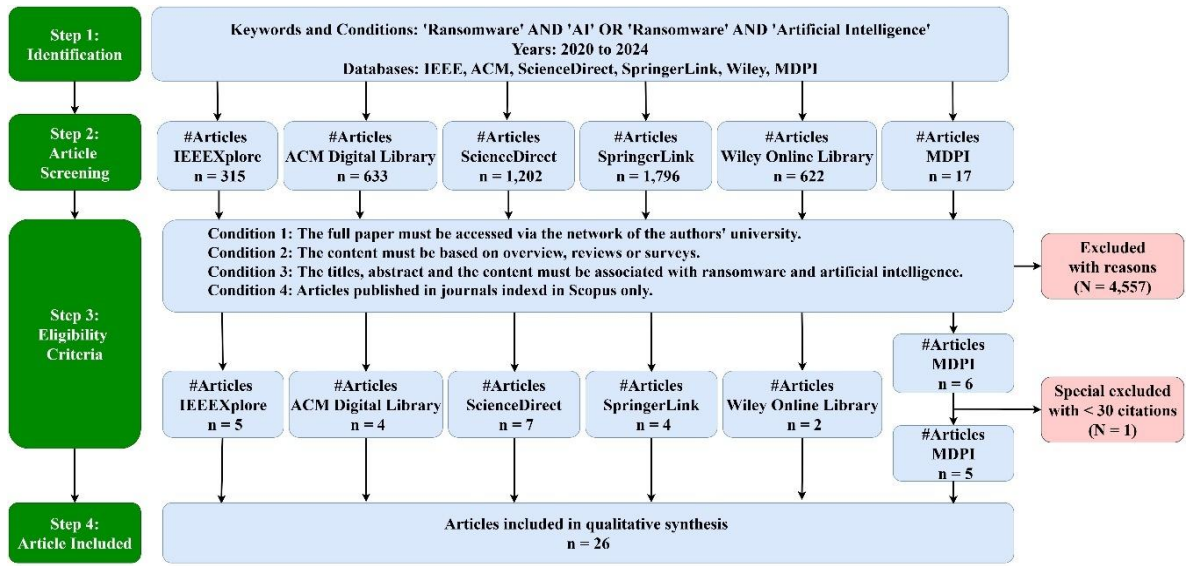

Figure 2. The PRISMA diagram for article selection in this study

Focus was made on three main issues: 1) the roles of AI in ransomware detection and prevention, 2) AI in ransomware mitigation, 3) challenges, and 4) future research directions. It was found that each issue has a list of related articles, as shown in Table 1. All issues are described in detail in the subsequent sections, i.e., sections 4-7.

Table 1. AI in ransomware research trends during the past five years, 2021-2024

| Ref. | Author(s) | Year | Applications of AI | | | | Contributions / Findings |
|------|-----------|------|---------------------|---|---|---|--------------------------|
| | | | Detection & Prevention | Mitigation | Challenges | Research Direction | |
| [1] | M. Robles-Carrillo & P. García-Teodoro | 2022 | | | | ✓ | Explores ransomware from technical and legal perspectives, stressing interdisciplinary defense frameworks. |
| [2] | J. Ferdous et al. | 2024 | ✓ | ✓ | ✓ | ✓ | Covers AI techniques in ransomware detection and mitigation, stressing hybrid frameworks, scalability, and ethical considerations. |
| [10] | G. Kirubavathi et al. | 2024 | ✓ | ✓ | ✓ | | Reviews ransomware's impact on healthcare, stressing risks to patient safety, data, and operational continuity. |
| [16] | J. Ferdous et al. | 2023 | ✓ | | ✓ | | Analyzes modern malware attack vectors and corresponding defense mechanisms, stressing anomaly detection and adaptive strategies. |
| [17] | M. Cen et al. | 2024 | ✓ | ✓ | ✓ | ✓ | Surveys ransomware early detection methods, emphasizing static, dynamic, and hybrid strategies for pre-encryption identification. |
| [18] | C. Beaman et al. | 2021 | ✓ | | ✓ | | Summarizes recent advances, challenges, and highlights double extortion as an evolving ransomware trend. |
| [19] | H. Oz et al. | 2022 | ✓ | ✓ | | | Comprehensive taxonomy of ransomware evolution, attack lifecycle, and countermeasures across static, dynamic, and hybrid approaches. |
| [22] | A. Kapoor et al. | 2021 | ✓ | ✓ | | | Reviews detection and mitigation schemes, emphasizing honeypots, decoys, and hybrid defense strategies. |
| [26] | J. Ispahany et al. | 2024 | ✓ | ✓ | ✓ | ✓ | Reviews ML-based ransomware detection, emphasizing limitations in datasets, scalability, and generalization to unseen ransomware variants. |
| [27] | S. Razaulla et al. | 2023 | ✓ | ✓ | ✓ | | Provides taxonomy of ransomware evolution; 72.8% of research focuses on detection (70% ML-based), while prediction and adversarial ML remain underexplored. |
| [28] | A. Alqahtani & T. F. Sheldon | 2022 | ✓ | ✓ | | | Surveys crypto-ransomware detection techniques, with focus on entropy and encryption pattern analysis. |
| [34] | T. McIntosh et al. | 2022 | ✓ | | ✓ | | Reviews ransomware mitigation frameworks, emphasizing hybrid static-dynamic approaches and challenges in real-time adaptation. |
| [36] | K. Begovic et al. | 2023 | ✓ | ✓ | ✓ | | Reviews entropy-based cryptographic ransomware detection, evaluating strengths and weaknesses of encryption analysis. |
| [45] | A. Alraizza & A. Algarni | 2023 | ✓ | ✓ | ✓ | ✓ | Reviews ransomware mitigation in modern contexts, stressing research challenges and future directions with focus on frameworks integrating detection, avoidance, and resilience. |
| [49] | U. Urooj et al. | 2021 | | | ✓ | ✓ | Surveys ransomware detection pre- and post-encryption, categorizing ML/DL studies; highlights hybrid static–dynamic approaches and the need for improved real-time early detection. |
| [50] | J. Singh & J. Singh | 2021 | ✓ | ✓ | ✓ | | Reviews ML methods for detecting malware in executables, focusing on feature engineering and classification challenges. |
| [51] | R. Moussaileb et al. | 2022 | ✓ | ✓ | ✓ | | Provides taxonomy of Windows ransomware families and detection methods, highlighting classification gaps and comparative effectiveness of detection strategies. |
| [52] | R. Ali et al. | 2022 | ✓ | ✓ | | ✓ | Reviews DL techniques for malware detection, highlighting scalability and performance benefits over traditional methods. |
| [53] | M. G. Gopinath & C. S. Sethuraman | 2023 | ✓ | ✓ | | | Comprehensive survey of DL malware detection, covering CNNs, RNNs, GANs, and their application to ransomware detection. |
| [54] | M. Gazzan, & F. T. Sheldon | 2023 | ✓ | ✓ | | ✓ | Explores early ransomware detection in ICS, recommending multi-modal DL and transfer learning for prediction. |
| [55] | G. M. Gaber et al. | 2024 | ✓ | | ✓ | | Systematic review of AI-based malware detection, identifying strengths, weaknesses, and dataset limitations in ransomware research. |
| [56] | X. Deng et al. | 2024 | ✓ | ✓ | | ✓ | Introduces DRL using PE header features, enabling adaptive and early ransomware detection without execution. |
| [57] | W. J. C. Chew et al. | 2024 | | ✓ | ✓ | | Proposes system call analysis with ML for real-time ransomware detection, decreasing false positives and improving speed. |
| [58] | L. Manning & A. Kowalska | 2023 | ✓ | | | | Examines ransomware threats to food supply chains, emphasizing cascading disruptions and need for proactive defense. |
| [59] | M. Humayun et al. | 2021 | | | ✓ | | Surveys ransomware in IoT environments, highlighting device vulnerabilities, service disruption, and lightweight detection approaches. |
| [60] | P. Formosa et al. | 2021 | | | | ✓ | Applies ethical principles to cybersecurity, stressing transparency, accountability, and moral duty for secure practices. |
| [61] | D. Smith et al. | 2022 | | | | ✓ | Explores ML algorithms (RF, SVM, DT) and frameworks, evaluating performance trade-offs in ransomware detection. |

## 4. AI IN RANSOMWARE DETECTION AND PREVENTION

The adoption of AI, ML and DL algorithms has revolutionized detection approaches, facilitating more resilient and efficient responses to ransomware attacks. See Figure 3, which is described as follows:

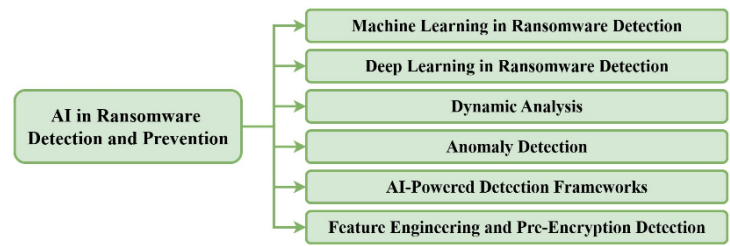

Figure 3. AI in ransomware detection and prevention

### 4.1 Machine Learning in Ransomware Detection

ML algorithms have shown strong results in classifying malware, including ransomware, due to their ability to handle its complexity and fast evolution [50]. These models analyze large datasets to spot patterns that distinguish ransomware from normal software. Their success depends on how well they extract key malware features using advanced analytical methods [50]. Support Vector Machines (SVM), Random Forest (RF), and Neural Networks (NNs) each have pros and cons [2]. SVM works well with well-designed features and in high-dimensional spaces, but it struggles with scalability and needs careful tuning [50]. RF offers very high accuracy in some studies, with few false positives [45]. In general, RF balances accuracy and interpretability, SVM suits smaller, structured datasets, and NN excels with complex, detailed input when resources allow. Hybrid and ensemble models that combine multiple algorithms are increasingly used to improve accuracy and robustness [51], [56]. ML-based systems often use both static and dynamic analysis to detect ransomware. Static analysis inspects executable files without running them, extracting features like opcode sequences and control flow graphs [55]. In contrast, dynamic analysis looks at behavior during execution, such as API calls, registry changes, and file modifications, helping to detect new ransomware variants [34], [51]. Behavioral features like API call sequences and runtime actions are key for training ML classifiers [19], [34]. API calls help apps exchange data, but ransomware often uses them to carry out harmful tasks [18]. By recognizing these patterns, ML models can tell if a program is safe or ransomware based on its behavior [18], [28]. Hybrid methods combining static and dynamic analysis have also shown promise in detecting ransomware before it starts encrypting data [52]. These techniques help ensure fast and accurate detection, reducing damage from attacks.

## 4.2 Deep Learning in Ransomware Detection

DL is increasingly used for ransomware detection due to its ability to learn complex patterns and handle large datasets well [52]. Common architectures include Convolutional Neural Networks (CNN) and Long Short-Term Memory (LSTM) [52]. These models detect malware by learning its behavioral patterns, like API calls and system interactions [53]. CNNs are good at processing structured data and spotting patterns, while LSTMs are effective in analyzing sequences such as API logs. Deep reinforcement learning is another promising method, especially for early detection. It can analyze static features like Portable Executable (PE) headers to identify ransomware without running it, reducing risks linked to dynamic analysis [56]. This is useful against ransomware that uses advanced evasion methods. RNN-based models also work well in analyzing API call sequences for efficient detection [53]. Generative Adversarial Networks (GANs) show promise in predicting ransomware variants. GANs can create new malware samples from known ones, helping researchers prepare for future threats [54]. LSTM and CNN models have also been reported to boost detection accuracy, reliability, and performance [27]. These models can adapt to new ransomware behaviors, keeping detection systems effective over time.

## 4.3 Dynamic Analysis

Dynamic analysis tracks activities during runtime, making it effective at spotting ransomware early in its execution [10], [54]. It examines behaviors like file encryption and registry changes, linking them to malicious actions [10], [17]. By simulating real environments, dynamic analysis shows how ransomware interacts with system resources. Event-based detection looks for signs of an attack, such as sudden CPU or disk use spikes, allowing early warning [28]. Combined with behavioral models, it uses API call patterns and process traces to tell ransomware apart from safe processes in real time [16]. This helps stop threats before much encryption happens [22], [58]. Tools like Pre-Encryption Detection Algorithm (PEDA) can be used to detect ransomware before encryption, preventing serious damage [22]. ML enhances these methods by spotting subtle irregularities, enabling proactive detection. Since dynamic analysis focuses on system changes, it can identify ransomware activities. Process Behavior Analysis (PBA) monitors system processes to catch malicious acts like unauthorized file encryption [10]. Behavioral models separate ransomware from safe processes by analyzing API calls and network logs [51]. NNs are widely used in advanced models to predict ransomware by learning from past data. This helps find small anomalies and speeds up response [26].

## 4.4 Anomaly Detection

Together, anomaly detection and behavioral analysis strengthen defenses by identifying unusual patterns, and when combined, these methods enhance scalability and adaptability for real-time ransomware detection and prevention [16], [45]. For example, DNS-based methods analyze traffic anomalies to contain ransomware within a single infected device [22]. When integrated with system behavior monitoring, this creates a strong defense that isolates threats without disrupting normal operations. Entropy-based monitoring further aids detection by distinguishing typical write operations from suspicious encryption, thereby reducing data loss and helping to contain ransomware early. Real-time I/O monitoring and entropy analysis enhance accuracy by detecting sudden malicious encryption changes. Hybrid analysis combines static and dynamic techniques, utilizing features from both to improve accuracy in ransomware detection [49]. Building on this, hybrid systems that integrate DL with behavioral analysis can accurately identify ransomware variants before significant harm occurs, and their dynamic detection and response capabilities make them well-suited for current cybersecurity strategies.

## 4.5 AI-Powered Detection Frameworks

AI-based detection frameworks use ML and DL to improve ransomware detection and prevention. These often categorize ransomware by its API call sequences and runtime actions [34]. Combined static and dynamic analysis methods are effective, offering more complete detection than single approaches [53]. For instance, hybrid models can find known ransomware signatures statically and detect new variants by their live behavior dynamically. Frameworks using supervised learning, like Decision Tree and Random Forest, are common for ransomware classification. These models show high accuracy and low false positives [45]. Also, AI-powered predictive models can combine operational data to forecast ransomware attacks more accurately [54]. These frameworks monitor network traffic and user behavior for proactive threat detection. Examples include SwiftR and SINNRD, which use hierarchical neural networks and spline interpolation for better ransomware detection [16]. They effectively expose hidden ransomware, lowering false positives, and surpass older methods against unknown threats [2], [16]. This shows the promise of advanced AI for real-time detection and prevention.

**4.6 Feature Engineering and Pre-Encryption Detection**

Effective ransomware detection depends on the quality of features used to train ML models. Common features include file and directory attributes (structural) and API calls (behavioral) [19]. Dynamic features, like execution logs, gathered during runtime offer strong detection because they are difficult to hide [26]. Feature engineering selects and improves these attributes for better performance and lower computing costs. Pre-encryption detection is vital for ransomware prevention. Catching ransomware activity before encryption allows stopping attacks before permanent damage [36]. Event-based detection, which identifies specific ransomware behaviors, is especially helpful here [28]. These behaviors include quick file extension changes, rapid file creation or modification, or unauthorized access to sensitive folders. Classifying ransomware families using ML with N-gram opcodes has also boosted detection accuracy, helping researchers categorize variants [10]. Integrating AI models that generalize beyond specific features improves the ability to detect new ransomware [55]. This adaptability keeps detection systems strong against evolving ransomware.

## 5. AI IN RANSOMWARE MITIGATION

Artificial intelligence (AI) plays a pivotal role in ransomware mitigation The topics of AI in Ransomware mitigation are shown in Figure 4.

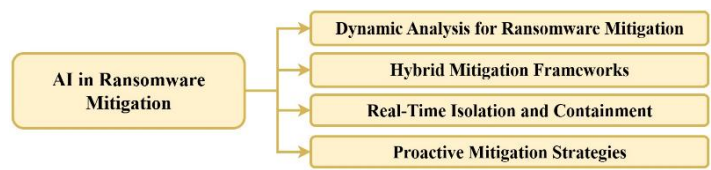

Figure 4. AI in ransomware mitigation

**5.1 Dynamic Analysis for Ransomware Mitigation**

Dynamic analysis captures real-time activities such as file operations, registry changes, process execution, and network activity [50]. This monitoring helps reveal ransomware behavior during execution, aiding in the detection of new variants [53]. It can also identify repetitive read/write patterns typical of ransomware, allowing early intervention [19]. Tools with dynamic analysis simulate system environments to observe interactions with system components. Sandboxing isolates malicious software and enables researchers to study targeted files and folders, encryption methods, and attempts to connect with command and control servers [52], [54]. These insights are valuable for designing effective defenses. Hybrid techniques that combine dynamic and static analysis overcome limitations of traditional approaches [50]. Such systems monitor encryption in progress and can terminate ransomware processes [36]. Dynamic detection therefore offers a key opportunity for containment [45]. Additional methods include honeypots and decoy files that attract ransomware and block execution [28], as well as dynamic blacklisting and I/O monitoring to detect malicious encryption through entropy analysis [51]. One good example of a hybrid framework, using exploit prevention [22]. Its Crypto Guard feature can recover encrypted files, acting as a proactive defense [22]. Backup strategies and checkpointing are also key for mitigation, though they may require lots of storage [22]. Some hybrid frameworks also use ML to improve detection. Analyzing static signatures and dynamic behaviors enhances robustness and scalability, making them better at adapting to new ransomware variants [45].

**5.3 Real-Time Isolation and Containment**

AI models can quickly isolate and contain ransomware attacks. By looking at operational and situational data, they predict threats and respond fast [54]. Situational awareness frameworks help organizations spot and isolate ransomware in real-time, causing little disruption to key systems [54]. Anomaly detection models use ML to find suspicious actions, like unauthorized file access. This allows immediate isolation of affected systems [17]. Real-time system call processing gives timely detection and allows quick containment [57]. For example, live network monitoring stops ransomware talking to command-and-control (C&C) servers, preventing further spread [17]. Hybrid detection systems effectively stop ransomware during encryption. By mixing static and dynamic analysis, they find and counter threats before major harm occurs [2]. Real-time responses within these systems lessen losses and risks by quickly detecting attacks [56]. Incident response layers in AI models improve containment. They isolate infected systems and cut off outside communication, ensuring ransomware does not spread to other devices or networks [10].

**5.4 Proactive Mitigation Strategies**

Prevention and protection against ransomware require a proactive approach, given the irreversible damage it can inflict on users' operations and data [54]. Proactive mitigation emphasizes early detection and prevention to reduce harm before encryption occurs. Static analysis identifies ransomware by examining executable files without running them, making detection safer and faster [56]. Maintaining effective models also requires proper labeling and continuous policy updates [56]. AI enables predictive analysis, helping organizations assess weaknesses, prioritize risks, and design proactive strategies [26]. Additional defense layers, such as blockchain-based response frameworks, protect critical data and contain ransomware spread, especially in sensitive sectors. Backup strategies combined with detection provide added protection and ensure recovery after an attack [27]. Real-time detection models that isolate suspicious activity before encryption further reduce potential damage [2]. Organizational practices, including recovery plans and system designs that segment and secure critical locations, enhance resilience [58]. Proactive approaches also include network segmentation, intrusion detection, endpoint protection, and continuous monitoring of user behavior and system activity to ensure rapid detection and response [54].

## 6. CHALLENGES IN RANSOMWARE DEFENSE AND RESEARCH USING AI

The challenges in combating ransomware with AI are categorized into six topics as shown in Figure 5, as follows:

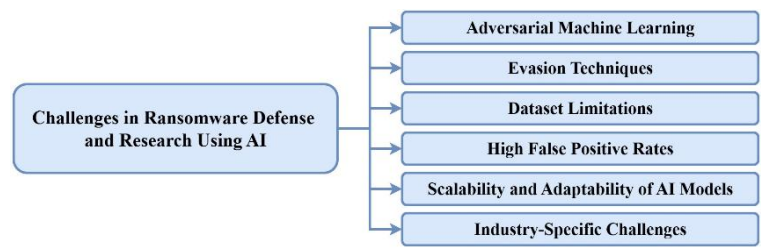

Figure 5. Challenges in Ransomware Defense Using AI

### 6.1 Adversarial Machine Learning

Adversarial Machine Learning (AML) refers to cyberattacks that manipulate ML decision-making processes by providing misleading inputs [27]. It also involves designing models capable of defending against such malicious attacks to enhance resilience and reliability, ensuring the safe and effective application of ML-based models in critical systems [27]. AML has become a significant concern in ML-based detection systems, as attackers can exploit vulnerabilities by generating malicious inputs, such as ransomware samples, to bypass detection [49], [50]. These intelligent evasion attacks gather information about detection models and then launch targeted attacks based on system and model vulnerabilities [49], [50]. One example is an adversarial evasion technique that successfully defeated a DL-based malware detection model, reducing detection accuracy from 94% to 50% when tested with modified adversarial samples [2]. Such attacks highlight the urgent need to develop robust ML models to counter adversarial threats and maintain trust in AI-driven detection systems [2]. The evolving nature of AML underscores its critical role in the ongoing arms race between cybersecurity defenses and malicious actors.

### 6.2 Evasion Techniques

Modern ransomware used advanced techniques to evade detection. Traditionally, methods (e.g., code encryption and polymorphism) were used to obscure ransomware files [26]. The recent development of fileless ransomware, which utilizes PowerShell scripts to break security and to download its payloads, now represents an evolution in these evasion strategies [26], [34]. Fileless malware bypasses executable file inspection, showcasing ransomware developers' ever evolving evasion tactics [34]. Advanced polymorphic and metamorphic ransomware make these threats virtually undetectable, rendering traditional detection mechanisms ineffective [51]. Moreover, adversarial evasion techniques have defeated DL models, reducing detection accuracy from 94% to 50% when tested with adversarial malware samples [2]. Evasive malware also resists analysis by commonly used tools (e.g., debuggers, decompilers, sandboxes, and Dynamic Binary Instrumentation (DBI) [55]. These challenges intensify as ransomware leverages obfuscation and evasion techniques that can operate for a very short duration to evade detection entirely [49]. Therefore, a novel detection mechanism that can counteract progressively stealthier and more adaptive threats is necessary.

### 6.3 Dataset Limitations

Using public and private datasets for ransomware detection presents challenges, including legal, proprietary, administrative, and ethical concerns [50]. Building public datasets with both safe and malicious files is difficult due to copyright restrictions and the risks of handling real malware samples [55]. Consequently, researchers usually rely on synthetic or sandbox-generated datasets, which fail to capture the full complexity of real-world attacks, limiting their usefulness [45], [55]. Many victims also hesitate to report incidents, resulting in a shortage of real-world data [45]. The absence of standardized datasets further complicates comparisons, as platforms like VirusShare and VirusTotal lack consistent creation methods and classification rules [17]. Small, private, and imbalanced datasets reduce reliability and generalization [55]. The shortage of varied, real-world data creates significant obstacles, especially since modern detection methods require large datasets to reduce false positives and adapt to new ransomware types [16]. Furthermore, transfer learning can also help to compensate data insufficiency by applying cross-domain knowledge [54].

## 6.4 High False Positive Rates

Achieving high detection rates while minimizing false-positive rates remains a significant challenge for anti-ransomware implementations. For example, streaming approaches by using finite state machines (FSMs) had promising detection capabilities [57]. However, the inclusion of a second FSM layer may marginally increase the false positive rate, highlighting the trade-offs involved in refining detection systems [57]. Heuristic and anomaly-based approaches offer broader detection capabilities, however, they are prone to false positive. In addition, they are resource-intensive. These factors have so far complicated their practical utilization. False positives or false alarms are incidents where legitimate software is wrongly specified as ransomware. They frustrate administrators and undermine the usability of detection systems [18]. Thus far, many existing systems struggle to balance between high detection accuracy and low false alarm rates. Unreliable anti-instrumentation techniques further contribute to misclassification issues [55]. Overcoming these challenges demands improvements in model design and evaluation to achieve robust performance in real-world environments while maintaining trust and efficiency [18], [34].

## 6.5 Scalability and Adaptability of AI Models

Transfer learning has shown promise in tackling scalability issues by reusing patterns from one ransomware type to detect others [2]. Still, scaling AI models is difficult due to the need to manage large datasets efficiently. Advanced techniques (e.g., Principal Component Analysis (PCA)) support select key features without sacrificing accuracy [2]. Small, unbalanced datasets with poor-quality features worsen the problem, as models may perform well on limited data but fail with new ransomware types [55]. Building public repositories that mix safe and malicious files could improve scalability, though legal and security barriers remain [55]. ML and adaptive methods still require testing to ensure scalability, reliability, and speed [16]. As DL has emerged as a promising technology for forecasting cyber-attacks using ransomware, it can be applied to develop DNNs capable of analyzing vast data from multiple sources and identifying patterns that signal attack likelihood [54]. Scalability is crucial for moving from academic studies to real-world deployments, where industry systems generate massive volumes of data [36]. Research aimed at faster and more efficient models will be vital for building robust AI tools that can handle evolving ransomware threats [16], [36].

## 6.6 Industry-Specific Challenges

Ransomware poses significant challenges across various societal sectors. More specifically, professional services, government, and healthcare account for 35%, 33%, and 28% of all cyber-attacks, respectively [36]. The healthcare industry is particularly vulnerable due to the critical and sensitive nature of the data involved. These data include patient information, medical records and operational systems, which are all highly valuable to cybercriminals [10]. The disruptions caused by ransomware in healthcare can indeed have devastating effects, such as compromised patient care, safety risks, and substantial financial losses [10]. Beyond healthcare, other industries also face systemic risks, including the food supply chain, where disruption of the provision of agricultural inputs has cascading effects [58]. The IoT further increases vulnerability because IoT devices usually lack necessary security measures. Attacks on IoT devices may disrupt critical services, such as public safety systems, leading to severe and irreversible consequences. The rapid adoption of IoT in various sectors has introduced new challenges, including increased susceptibility to ransomware attacks. These challenges emphasize the need for more robust cybersecurity measures [59], requiring focused strategies to strengthen defenses and mitigate the negative impacts of cyber-attacks using ransomware.

## 7. FUTURE RANSOMWARE RESEARCH DIRECTIONS ASSOCIATED WITH AI

For future research directions toward AI-associated ransomware detection, prevention and mitigation, there are eight topics as shown in Figure 6, which are described as follows:

## 7.1 Hybrid AI Frameworks

Hybrid methods for ransomware detection have emerged as critical tools for addressing the limitations inherent in static and dynamic analysis techniques. Hybrid analysis combines the features and strengths of both approaches to achieve the most precise detection results [49]. Examples such as SwiftR, RansHunt, and RansomWall exemplify their effectiveness, achieving high accuracy in ransomware identification by integrating diverse analysis techniques [2]. A notable implementation of this approach is RANDS, a Windows-based anti-ransomware application that uses a multi-tier framework, combining ML algorithms to classify ransomware families and variants [61]. The hybrid model integrates Decision Tree and Naïve Bayes algorithms, yielding an average detection accuracy of approximately 96.3% with an error rate of about 1.3% in real-time assessments [61]. These advancements underline the vital role of hybrid methods in enhancing ransomware detection capabilities in dynamic cybersecurity realms.

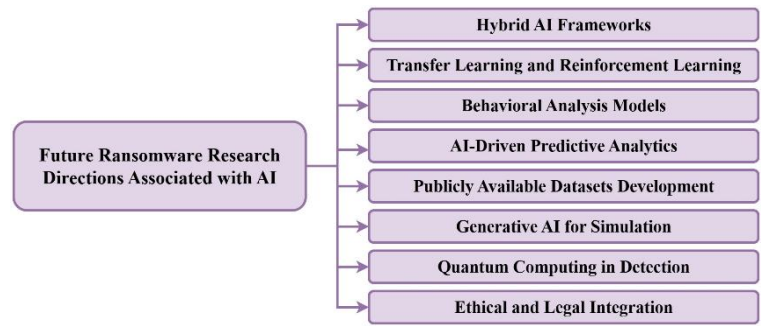

Figure 6. Ransomware research directions associated with AI in the future

## 7.2 Transfer Learning and Reinforcement Learning

Researchers have adopted advanced learning techniques to enhance ransomware detection. Transfer learning reduces computational demands by leveraging pre-trained models for effective detection [2]. Besides, research can emphasize developing transfer-learning models which can be applied by training on one domain and then applied to another domain to gain up the accuracy of forecastings in new applications [54]. A DRL-based method has also been proposed for malware detection, where binaries are mapped into integer ranges for visualization, followed by DL-based classification. This approach outperformed its counterparts, demonstrating DL's superior capability in malware detection [52]. DRL further combines deep and reinforcement learning to provide adaptive, data-driven, real-time ransomware detection [2]. In one framework, static features from ransomware executables were extracted to serve as the state for DRL models [56]. These methods adapt to new threats by identifying patterns in ransomware samples and network logs, enhancing detection accuracy [56]. Trained on extensive datasets, DRL achieves high precision in capturing ransomware behaviors, enabling rapid detection and effective responses to mitigate potential losses [56].

## 7.3 Behavioral Analysis Models

Behavioral analysis is a critical approach in ransomware detection. It focuses on monitoring or observing the behavior of running processes and disclosing anomalies indicative of malicious activity [45]. This methodology applies feature selection of logged data and extracting the most beneficial information, enabling the identification of malicious attacks [45]. For instance, a version of behavioral detection methodologies utilized an ML baseline model to simulate and forecast user behavior patterns at the micro level, identifying potential vulnerabilities [45]. Real-time behavioral analysis is generally preferred, when promptly detecting and responding to cyber-attacks are needed. However, dynamic analysis, often conducted in controlled environments, is also necessary for identifying new and previously unseen ransomware variants, which can enhance real-time behavioral analysis [17], [45]. These advanced techniques, including realistic simulations of evasive behaviors, provide a robust framework for countering sophisticated ransomware threats [56].

## 7.4 AI-Driven Predictive

AI-driven predictive approaches play a crucial role in ransomware detection and prevention by enabling early identification of malicious activities [17]. One research direction is to develop multi-modal DL models that can consider several data sources (e.g., user behavior, network traffic and system logs) to make more accurate predictions [54]. Several studies have also explored ransomware threats and proposed detection and prevention solutions to mitigate such attacks [17]. Advanced frameworks leverage DRL to learn ransomware features through interactions between an agent and its environment, classifying samples as ransomware or benign [56]. In addition, lightweight networks that use the portable executable (PE) header from some byte files enhance detection speed while avoiding the need to execute ransomware samples [56]. These innovations support rapid and effective detection, providing strong defenses against evolving ransomware threats while minimizing risks to systems and data.

**7.5 Publicly Available Datasets Development**

The development and use of publicly available datasets are crucial for advancing ransomware detection analytics. A significant challenge in this field is the lack of such datasets, which are essential for training and evaluating ma-chine learning models [26]. Comprehensive studies have provided valuable insights into dataset collection and sources, supporting ransomware detection across diverse platforms [49]. These studies also present a range of datasets and analysis tools used in dynamic analysis, enabling the training and testing of ML-based and DL-based ransomware detection systems [49]. Experiments on real-world datasets show that unseen ransomware can be rapidly detected by the proposed framework. These findings highlight the importance of robust dataset development for advancing detection capabilities [56].

**7.6 Generative AI for Simulation**

Generative adversarial networks (GANs) have shown a pivotal role in advancing ransomware research, particularly in generating artificial ransomware examples that mimic potential future attack behaviors [54]. GANs are designed to create new sets of data instances resembling existing datasets, making them highly valuable for adversarial learning tasks, especially in scenarios where data on zero-day attacks is limited [26]. These networks have been employed to generate synthetic datasets that augment real attack patterns. They address the evolving nature of ransomware and its tactics in zero-day scenarios. Using GANs to simulate ransomware attacks offers critical insights into potential vulnerabilities and also supports the development of effective defense strategies. This approach can be developed for testing environments, allowing researchers to evaluate countermeasures without the risks from "live" ransomware.

**7.7 Quantum Computing in Ransomware**

Quantum computing (QC) is a promising technology, which has very high potential to revolutionize ransomware detection and mitigation. Recent studies suggest that quantum computing will reach maturity in the near future, allowing detection algorithms to be enriched with quantum capabilities that expedite the identification and decryption of encrypted malware [61]. Moreover, defenders could employ quantum systems for proactive monitoring, enabling the decryption of malicious communications and thereby disrupting ransomware operations [61]. Leveraging these capabilities, QC can also interfere with ransomware attack sequences even after files have been encrypted [61]. Collectively, these advancements highlight the transformative role of quantum computing in strengthening cybersecurity defenses against ransomware.

**7.8 Ethical and Legal Integration**

Fighting ransomware requires strong ethical, legal, and governance frameworks. Ethically, users must be protected, and fairness in AI-based detection must be ensured. Legally, aligning international laws can help close gaps exploited by cybercriminals. Cybersecurity systems should also be explainable and transparent, with individuals and organizations held accountable for their operation [60]. Strengthening these areas enhances overall resilience. Both technical and legal perspectives are vital to address the problem [1]. Since ransomware presents technical, social, political, and legal challenges, laws must counter such threats while maintaining public trust [1]. System operators are legally obliged to follow best practices for protecting and backing up user data, reflecting their duty of fairness and transparency. Organizations likewise bear an ethical responsibility to invest in strong security measures, such as backup and recovery. Ultimately, integrating technological and legal approaches is essential for effective prevention and response [1].

**8.    DISCUSSION**

This systematic review is not only a summary of previous reviews, it goes beyond those works, offering new findings on how AI is changing ransomware research and defense [2], [55]. Using the PRISMA method, a clear selection of articles was ensured, making the findings credible and providing a full overview of AI's role in ransomware detection, prevention, and mitigation [2]. AI technologies, especially ML, DL, and hybrid models, have greatly improved detection. They combine static and dynamic analysis with anomaly detection to spot ransomware before encryption. These methods are vital for isolating unusual processes and stopping attacks before major harm [ 2], [17]. AI-based defense focuses on real-time isolation, hybrid plans, and quick responses to lessen attack impact and help systems recover [34]. However, several challenges remain. Adversarial machine learning can compromise detection models and bypass existing defenses. Incomplete or imbalanced datasets negatively affect model training, generalizability, and robustness, limiting the development of reliable detection systems. High false-positive rates and scalability constraints further hinder real-world deployment, particularly in critical sectors such as finance and healthcare that require context-aware and highly dependable solutions [26], [55].

Technical skills alone are not enough against smarter ransomware. Ethical, legal, and governance factors must be incorporated to ensure AI tools operate responsibly and comply with regulations and privacy standards [1]. Future research should build hybrid AI models, use transfer and reinforcement learning, create standard public datasets, employ generative models like GANs, and explore quantum computing [2], [55]. This study also shows the constant interaction between new ransomware tactics and AI defense, pointing out key areas for more study. The findings of this study act as a guide for future work and stress the need for cooperation among researchers, industry, and lawmakers to create strong, ethical, and technically sound ways to fight ransomware globally [1], [34].

While this review offers a comprehensive synthesis, certain limitations must be acknowledged. Many referenced studies rely on synthetic or sandbox-generated datasets, which may not reflect the full complexity of real-world ransomware behavior. Public repositories often lack standardized labeling, leading to inconsistencies in model evaluation [26], [55]. These factors introduce potential bias and reduce the generalizability of findings to novel ransomware variants [2]. Additionally, legal and ethical barriers limit access to diverse real-world data, hindering reproducibility and robust validation of AI-based detection methods in practical settings [1].

## 9. CONCLUSION

This comprehensive study synthesizes research conducted over the past five years to explore the integration of AI, particularly ML and DL, in ransomware detection, prevention, and mitigation. Unlike typical reviews, it adopts a "review of reviews" approach guided by the PRISMA concept. This method consolidates findings from various sources, providing a clear, comprehensive view of ransomware threats and how AI is evolving to counter them systematically. Key insights include the strategic use of hybrid models, which combine both static and dynamic analyses, for more proactive defense. The study highlights techniques such as behavioral anomaly detection and pre-encryption monitoring, which help identify ransomware early and respond effectively to increasingly sophisticated attacks. The review also tracks ransomware's evolution from older forms to current ones such as Ransomware-as-a-Service (RaaS) and double extortion, which have greatly increased both impact and complexity. Despite technical advances, challenges remain. These include the need for scalable AI models, difficulties in obtaining diverse and high-quality datasets, and the rise of adversarial techniques that can evade detection. Addressing these issues requires ongoing improvements in AI design and data practices. Finally, the study underscores the importance of fostering collaboration among academia, industry, and policy sectors to close the gap between research and implementation. This paper offers both a consolidated foundation and a forward-looking guide, supporting the development of responsible, resilient, and adaptable cybersecurity strategies to safeguard critical infrastructures and digital ecosystems worldwide.

## ACKNOWLEDGEMENTS

This research has been facilitated by Rajamangala University of Technology Phra Nakhon. Thank you Dr.Tharis Thimthong, Dr.Prateep Pholchanngam and Assoc. Prof. Dr.Pramuk Unahalekhaka for supporting.

## FUNDING INFORMATION

This research has received funding support from the NSRF via the Program Management Unit for Human Resources & Institutional Development, Research and Innovation (PMU-B) [grant number B13F660061]. It also received research support from The Brain Stem Co., Ltd.

## AUTHOR CONTRIBUTIONS STATEMENT

Following the Contributor Roles Taxonomy (CRediT) to recognize individual author contributions, reduce authorship disputes, and facilitate collaboration, each author contribute roles as follows:

| Name of Author | C | M | So | Va | Fo | I | R | D | O | E | Vi | Su | P | Fu |
|---|---|---|---|---|---|---|---|---|---|---|---|---|---|---|
| Therdpong Daengsi | ✓ | ✓ | ✓ | ✓ | ✓ | ✓ | ✓ | ✓ | ✓ | ✓ | ✓ | ✓ | ✓ | ✓ |
| Phisit Pornpongtechavanich | | | ✓ | ✓ | | ✓ | | | | ✓ | ✓ | | ✓ | |
| Paradorn Boonpoor | ✓ | ✓ | | | | | | | | ✓ | | | | |
| Kathawut Wattanachukul | ✓ | ✓ | | | | | | | | ✓ | | | | |
| Korn Puangnak | ✓ | | | | | | ✓ | | | ✓ | | ✓ | ✓ | ✓ |

| | C | M | So | Va | Fo | I | R | D | O | E | Vi | Su | P | Fu |
|---|---|---|---|---|---|---|---|---|---|---|---|---|---|---|
| Kritphon Phanrattanachai | | ✓ | | | ✓ | | ✓ | | ✓ | | | ✓ | ✓ | ✓ |
| Pongpisit Wuttidittachotti | ✓ | | | | | | | | | | | | | |
| Paramate Horkaew | | ✓ | | | | | | | ✓ | | ✓ | | | |

C  : **C**onceptualization
M  : **M**ethodology
So  : **So**ftware
Va  : **Va**lidation

Fo  : **Fo**rmal analysis
I  : **I**nvestigation
R  : **R**esources
D  : **D**ata Curation

O  : Writing - **O**riginal Draft
E  : Writing - Review & **E**diting
Vi  : **Vi**sualization

Su  : **Su**pervision
P  : **P**roject administration
Fu  : **Fu**nding acquisition

## CONFLICT OF INTEREST STATEMENT

The authors declare no conflict of interest.

## DATA AVAILABILITY

The data that support the findings of this study are available from the first author, [TD], or the corresponding author, [PP], upon reasonable request.

## DECLARATION OF GENERATIVE AI IN WRITING PROCESS

The authors generated the content, ideas, and findings presented in the manuscript. ChatGPT and other generative AI were then used to assist with proofreading, improving, and refining the English language. Finally, the authors reviewed and validated the final version to ensure its accuracy and integrity prior to submission.

## BIOGRAPHIES OF AUTHORS

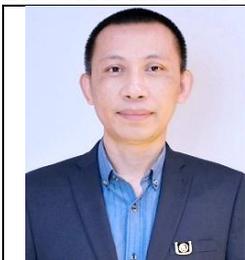

**Therdpong Daengsi** 🔴 🔀 ⊠ ⊙  Therdpong Daengsi is an Assistant Professor and currently serves as Head of the Sustainable Industrial Management Engineering Department in the Faculty of Engineering at Rajamangala University of Technology Phra Nakhon (RMUTP). He received his B.Eng. in Electrical Engineering from King Mongkut's University of Technology North Bangkok (KMUTNB) in 1997. He later earned an M.Sc. in Information and Communication Technology in 2008, from Assumption University and a Ph.D. in Information Technology at KMUTNB in 2012. With 19 years of experience in the telecommunications sector, he has also worked as an independent academic for a short period before becoming a lecturer. His research interests include QoS/QoE, mobile networks, multimedia communications, cybersecurity, and artificial intelligence. He can be contacted via email: therdpong.d@rmutp.ac.th.

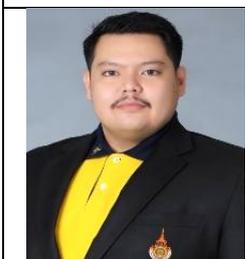

**Phisit Pornpongtechavanich** 🔴 🔀 ⊠ ⊙ is an Assistant Professor in the Faculty of Industry and Technology, Rajamangala University of Technology Rattanakosin, Wang Klai Kangwon Campus (RMUTR_KKW). Also, he currently serves as the Deputy Dean for Academic Affairs and Curriculum and as the Head of the Department of Information Technology and Digital Innovation. In 2012, he received his Bachelor of technology in information technology from RMUTR_KKW. He obtained a scholarship and then received a Master of Science in information technology from KMUTNB in 2014 and a Ph.D. in Information and Communication Technology for Education in 2023. His research interests include security, Deep Learning, AI, IoT, VoIP quality measurement, QoE/QoS, mobile networks, and multimedia communications. He can be contacted via email: phisit.kha@rmutr.ac.th.

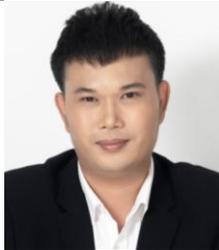 **Paradorn Boonpoor**    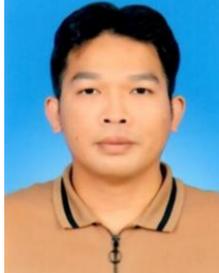 is currently a Forestry Officer at the Department of National Parks, Wildlife and Plant Conservation. He received a B.Sc in Forestry (first class honors) from Kasetsart University in 2015. Then, he received a M.Sc. in Business Analytics and Data Science from National Institute of Development Administration in 2021. He previously worked in the field of data analytics and data sciences. He was a Post-master Researcher at RMUTP, which received funding support from the NSRF via the Program Management Unit for Human Resources & Institutional Development, Research and Innovation (PMU-B). His research interests include Data Science, ML, AI, NLP and Computer Vision. He can be contacted via email: paradorn.sp@rmutp.ac.th.

**Kathawut Wattanachukul**   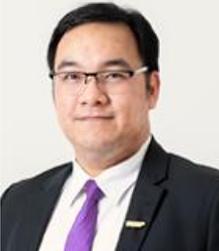 is now a PhD candidate in Network and Information Security Management. He obtained Bachelor of Engineering in Electrical Engineering from SWU and Master of Science (Information Technology Management) from NIDA in 2008 and 2019, respectively. He was also a Post-master Researcher at RMUTP, which received funding support from the NSRF via the Program Management Unit for Human Resources & Institutional Development, Research and Innovation (PMU-B). His has more than 20 years of experience in IT covering telecommunications, networking, systems, cybersecurity, cloud, data center, IoT, etc, as a Network and System Engineer for technology service providers and services of technology business consulting firms. He can be contacted via email: kathawut.sp@rmutp.ac.th.

**Korn Puangnak**   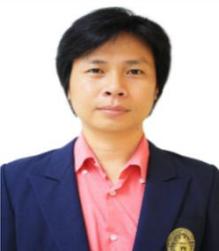 is an Assistant Professor in the Faculty of Engineering, Rajamangala University of Technology Phra Nakhon (RMUTP), Thailand. He received B.Eng. and M.Eng. in computer engineering from King Mongkut's Institute of Technology Ladkrabang (KMITL), Thailand, in 2006 and April 2011, respectively. He received D.Eng. in Sustainable Industrial Management Engineering from RMUTP in 2022. At present, he is a Vice President in RMUTP. His main research interests include machine learning, image processing, computer vision and intelligent transport system (ITS). He can be contacted via email: korn.p@rmutp.ac.th.

**Kritphon Phanrattanachai**   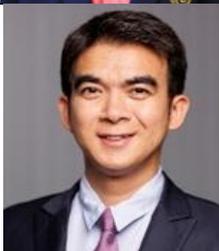 is an Assistant Professor at the Faculty of Agricultural and Industrial Technology, Phetchabun Rajabhat University (PCRU), Thailand. He received a Bachelor of Science degree in Electrical Industry from Phetchabun Rajabhat University in 2002, a Master's degree in Electrical Technology from KMUTNB in 2009, and a PhD degree in Information Technology from KMUTNB in November 2019. He is currently the Director of the PCRU Research and Development Institute. His research interests include electrical circuit synthesis, simulation of linear and non-linear circuits and systems, Internet of Things (IoT), Quality of Service (QoS/QoE), mobile networks, and telecommunications. He can be contacted at kritphon.ai@pcru.ac.th.

**Pongpisit Wuttidittachotti**   is currently the Thailand Cyber Security & Privacy Officer (CSPO) at Huawei Technologies (Thailand) Co., Ltd. Previously, he was an Associate Professor and the Head of the Department of Digital Network and Information Security Management (DNS) at the Faculty of Information Technology and Digital Innovation, KMUTNB, Thailand. His research interests include Cybersecurity, Management Systems, VoIP, QoS/QoE, mobile and wireless networks, multimedia communications, telecommunications, data science, IoT, and AI. He has been a member of the ISACA Bangkok Chapter since 2017. He also serves as a committee member for the Thailand Information Security Association (TISA) and the ISC2 Bangkok Chapter. In 2022, he received the National Excellent Lecturer in Social Service Award from The Council of University Faculty Senate of Thailand (CUFST). He can be contacted via email: pongpisit.w@itd.kmutnb.ac.th.

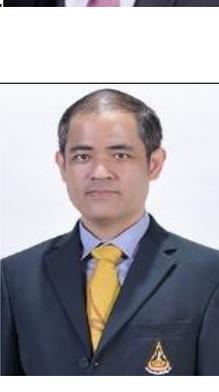 **Paramate Horkaew**    is an Associate Professor with the School of Computer Engineering, Suranaree University of Technology. His B.Eng. (Hons.) in Telecommunication Engineering (1st) (1999) from King Mongkut's Institute of Technology, Ladkrabang (KMITL). During his undergraduate study, he worked as an RA in medical informatics at the Computed Tomography Laboratory, NECTEC (1997-99). He had then continued his research, supported by the Ministry of Science, in medical image computing at the Visual Information Processing (VIP) group and the Royal Society/ Wolfson Foundation Medical Image Computing (MIC) laboratory, Imperial College London (2000-04). His main research interests include Networks, Remote Sensing, Computational Anatomy, Digital Geometry Processing, Computer Vision and Graphics. He can be contacted at email: phorkaew@sut.ac.th.